\documentclass{article}
\usepackage{spadre2008}
\usepackage{graphicx}
\frompage{000} \topage{000}                                              
\def\rightmark{Strange hadron(\ks, \lam and \xi) elliptic flow from 200GeV \cucu collisions}
\def\leftmark{S.S Shi}
\newcommand{\sqrtsNN}{\mbox{$\sqrt{\mathrm{s}_{_{\mathrm{NN}}}}$}}

\newcommand{\lam}{$\Lambda$ }
\newcommand{\ks}{$\mathrm{K}^{0}_{S}$ }

\def \auau  {$Au + Au$ }
\def \cucu  {$Cu + Cu$ }

\def \xi       {$\Xi$ }

\title{Strange hadron(\ks, \lam and \xi) elliptic flow from 200 GeV \cucu collisions}
\authors{
{Shusu Shi for the STAR collaboration %
}\\[2.812mm]
{\normalsize \hspace*{-8pt}$^1$ Nuclear Science Division, LBNL,
Berkeley 94720, USA\\[0.2ex]
\hspace*{-8pt}$^2$ Institute of Particle Physics, CCNU, Wuhan
430079, China
}}

\abstract{Collective flow reflects dynamical evolution in
high-energy heavy ion collisions. In particular, the strange hadron
elliptic flow reflects early collision dynamics~\cite{NPA757}. We
present results from a systematic analysis of the centrality
dependence of strange hadron elliptic flow ($v_2$) measurement of
$K_{S}^{0}$, $\Lambda$ and \xi for \cucu collisions at 200 GeV.
Results for \cucu collisions are compared with results previously
reported for \auau collisions. We will also compare our data with
results from ideal hydrodynamic calculations. } 
\begin{document}
\maketitle \setcounter{page}{1}
\section{Introduction}
The characterization of the elliptic flow of produced particles by
their azimuthal anisotropy has been proven to be one of the most
fruitful probes of the dynamics in \auau collisions at the
Relativistic Heavy Ion
Collider(RHIC)~\cite{flow2,flow3,flow4,flow5,flow6}. Study elliptic
flow in smaller collision systems, such as \cucu, which has
one-third nucleons in \auau, is beneficial. Because exactly how flow
scales with collision systems, such as system size, geometry,
constituent quarks , transverse momentum and transverse energy, is
crucial to the understanding of the properties of the produced
matter. Hydrodynamic model calculations, with the assumption of
ideal fluid behavior (no viscosity), have been successful when
compared with the experimental data at RHIC~\cite{flowPRC,starwp}.
In this proceeding, we extend the comparison with ideal hydrodynamic
calculations to different systems.
\begin{table}[ht]
\centering
\begin{tabular}{||c|c|c|c|c||}
\hline\hline
 &\multicolumn{3}{c|}{\cucu}  &\auau \\ \hline
                 &0-60\%   &0-20\% &20-60\% &0-80\% \\ \hline
 $dN_{ch}/d\eta$ &74 &132  &45  &225 \\ \hline
 $N_{part}$ &51  &87 &34 &126 \\ \hline
 $N_{bin}$ &80  &156 &43 & 293\\ \hline
$\varepsilon_{part}$ &$0.252$ &$0.184$  &$0.393$ &$0.214$ \\
\hline\hline
\end{tabular}\vspace*{-0.5cm}
\caption{List of $dN_{ch}/d\eta$, number of participants $N_{part}$,
number of binary collisions $N_{bin}$, and participant eccentricity
$\varepsilon_{part}$ for three centrality bins in 200 GeV \cucu
collisions and 0-80\% 200 GeV \auau collisions.} \label{tab:glauber}
\end{table}

\begin{figure}[ht]\centering\vspace*{0.cm}
\includegraphics[totalheight=0.5\textwidth]{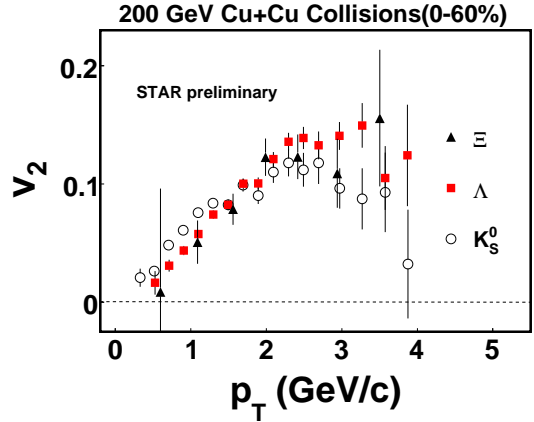}\vspace*{-0.5cm}
\caption{$v_2$ as a function of $p_T$ for \ks (open-circles), \lam
(filled-squares) and \xi (filled-triangles) in 0-60\% \cucu
collisions at \sqrtsNN = 200 GeV.} \label{fig1}
\end{figure}

\begin{figure}[htb]\centering
\vspace*{-1.cm}
\includegraphics[totalheight=0.5\textwidth]{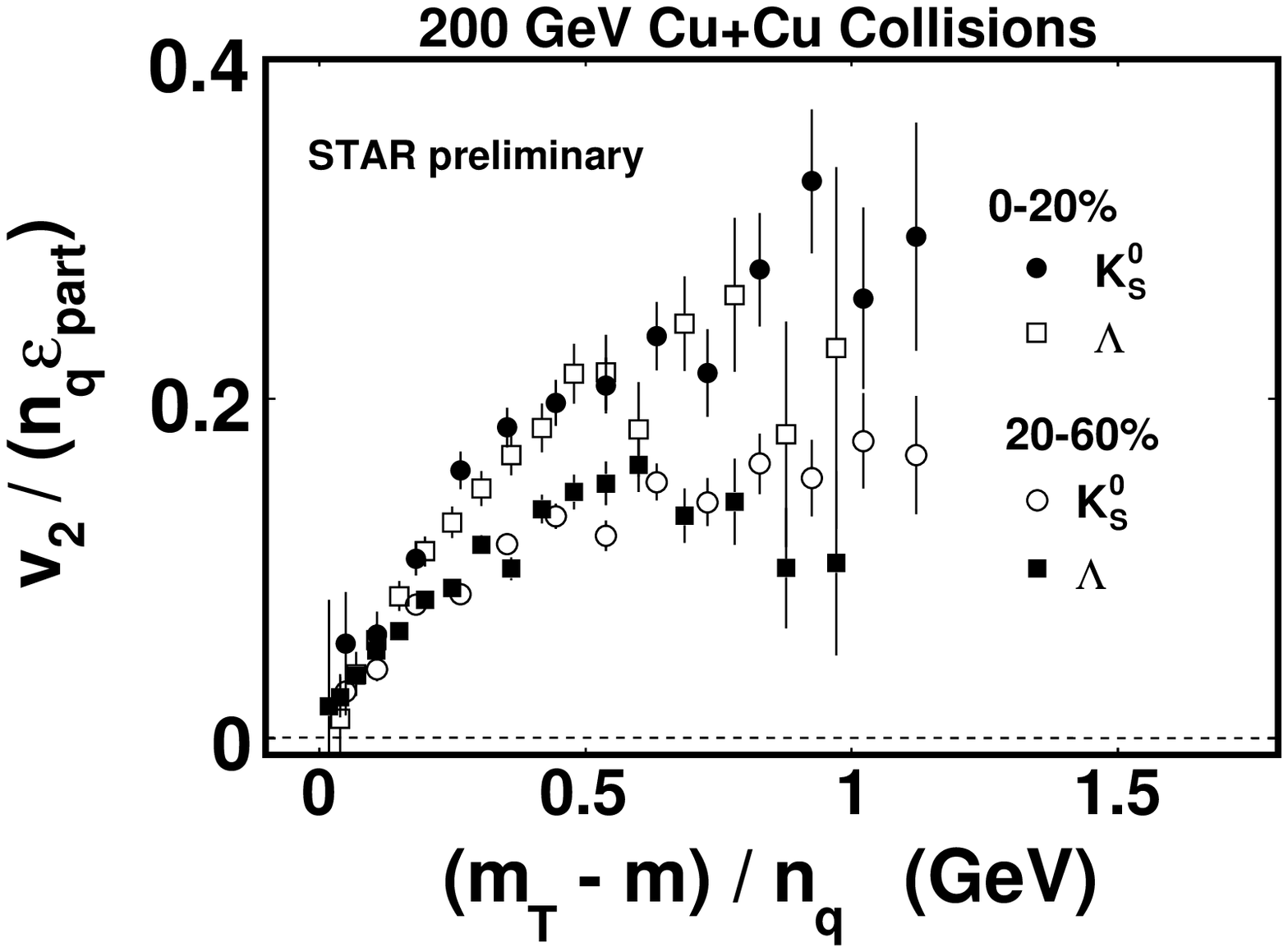}
\vspace*{-0.5cm} \caption{The eccentricity ($\varepsilon_{part}$)
and number of quark ($n_q$) scaled $v_2$ versus ($m_T - m)/n_q$ from
0-20\% (filled-circles: \ks, open-squares: \lam) and 20-60\%
(open-circles: \ks, filled-squares: \lam) \cucu collisions at
\sqrtsNN = 200 GeV.} \label{fig2}
\end{figure}

\begin{figure}[ht]\centering
\vspace*{-0.8cm}
\includegraphics[width=1.\textwidth]{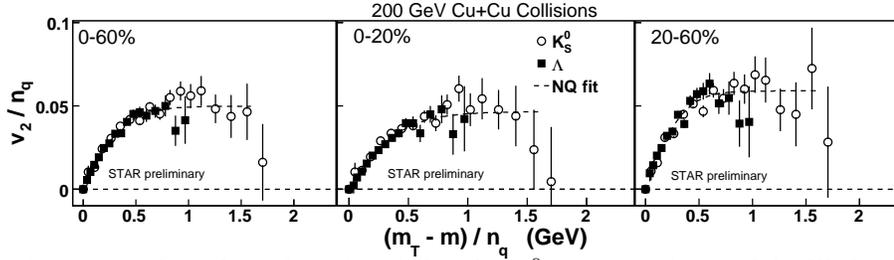}\vspace*{-0.5cm}
\caption{Number of quark ($n_q$) scaled $v_2$ for \ks (open-circles)
and \lam (filled-squares) as a function of ($m_T - m)/n_q$ from
three centrality bins. The results of the fits~\cite{NCQ} are shown
as dashed-lines in the figure. } \label{fig3}
\end{figure}

\begin{figure}[th]
\centering{
\includegraphics[totalheight=0.8\textwidth]
{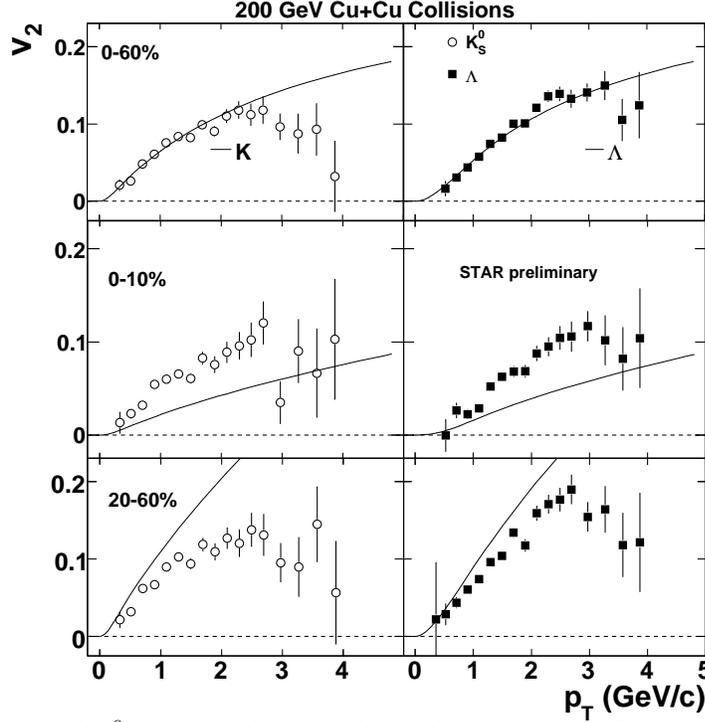}} \vspace*{-0.5cm}\caption{$v_2$ of \ks (open-circles)
and \lam (solid-squares) as a function of $p_T$ for three
centralities 0-60\%, 0-10\% and 20-60\% in \cucu collisions at
\sqrtsNN = 200 GeV. For comparison, ideal hydrodynamic calculations
are also shown as lines.} \label{fig4}
\end{figure}

\begin{figure}[ht]\centering{
\includegraphics[totalheight=0.5\textwidth]{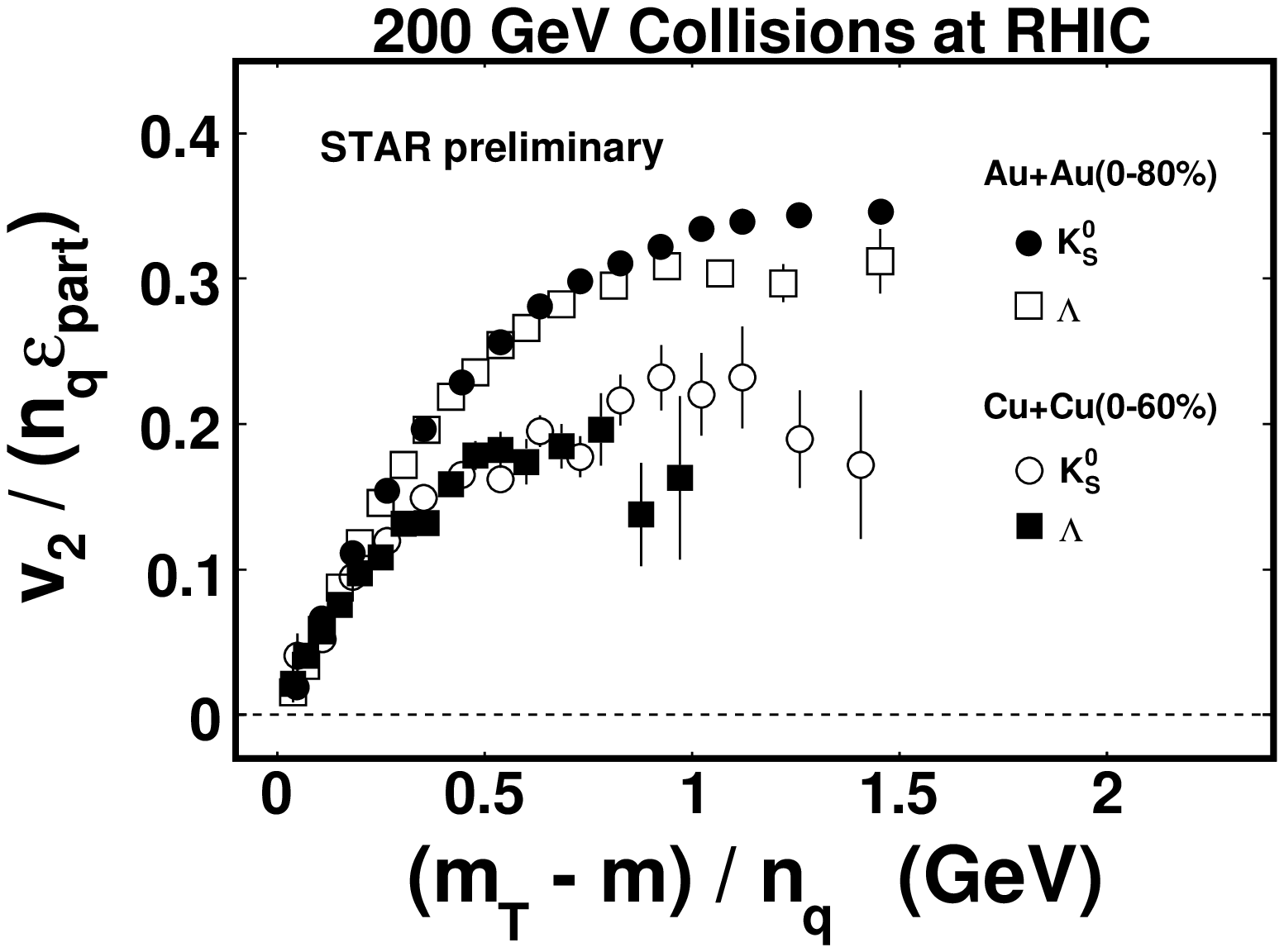}}\vspace*{-0.5cm}
\caption{The eccentricity ($\varepsilon_{part}$) and number of quark
($n_q$) scaled $v_2$ versus ($m_T - m)/n_q$ from 0-60\% \cucu
(open-circles: \ks, filled-squares: \lam) and 0-80\% \auau
collisions (filled-circles: \ks, open-squares: \lam) at \sqrtsNN =
200 GeV.} \label{fig5}
\end{figure}

\section{Methods and Analysis}
In this proceeding, we report results from \sqrtsNN = 200 GeV \cucu
collisions. Data were taken from Run 5 (2005). STAR's Time
Projection Chamber (TPC)~\cite{STARtpc} is used as the main detector
for particle identifications. The centrality was determined by the
number of tracks from the $|\eta| \le 0.5$. Two Forward Time
Projection Chambers (FTPCs) were also used for event plane
determinations.  FTPC has the coverage of 2.5 $\le |\eta \le$ 4, and
the pseudorapidity between FTPC and TPC allows us to reduce some of
the non-flow effects.

The PID is achieved via topologically reconstructed hadrons: \ks
$\rightarrow \pi^{+} + \pi^{-}$, \lam $\rightarrow p + \pi^{-}$
($\overline{\Lambda} \rightarrow \overline{p} + \pi^{+}$) and
 $\Xi^{-} \rightarrow$ \lam $+\ \pi^{-}$ ($\overline{\Xi}^{+}
\rightarrow$ $\overline{\Lambda}$+\ $\pi^{+}$). The detailed
description of the procedure can be found in
Refs.~\cite{klv2_130GeV,starklv2}.

 $v_2$ analysis was done in three centrality bins. The
corresponding number of charged hadrons, number of participants,
number of binary collisions, and the participant eccentricity for
each centrality bin are listed in
 Table I. For comparison, parameters for 0-80\% \auau collisions are also
 listed in the table.

The observed $v_{2}$ is the second harmonic of the azimuthal
distribution of particles with respect to this event plane:
\begin{equation} \label{v2obs} v_{2}^{obs}\ =\ \langle
\cos[2(\phi-\Psi_2)]\rangle
\end{equation}
where angle brackets denote an average over all particles with their
azimuthal angle $\phi$ in a given phase space. To take into account
the smearing of the estimate event plane around the true reaction
plane, the real $v_2$ has to be corrected for the event plane
resolution by
\begin{equation} \label{v2EP2} v_{2}\ =\
\frac{v_2^{obs}}{\langle \cos[2(\Psi_2-\Psi_r)]\rangle}
\end{equation}

For $v_2$ of the identified particles, \ks, \lam and \xi, the $v_2$
versus $m_{inv}$ method~\cite{v2minv} is used in this analysis. We
use $\Lambda$ (\xi) to denote $\Lambda +\overline{\Lambda}$ ($\Xi^-
+ \overline{\Xi}^+$) unless stated otherwise.

\section{Results}
Figure~\ref{fig1} shows $v_2$ as a function of $p_T$ for
\ks(open-circles), \lam(filled-squares) and \xi(filled-triangles) in
0-60\% \cucu collisions at \sqrtsNN = 200 GeV. At low $p_T$, \lam
$v_2$ is smaller than \ks; At high $p_T$, baryon(\lam, \xi) $v_2$ is
systematically greater than meson(\ks). \ks and \lam $v_2$ cross
over at $p_T$ 1.5 - 2.0GeV.

Figure~\ref{fig2} shows $n_q$-scaled $v_2$ normalized by participant
eccentricity as a function of ($m_T - m)/n_q$ for \ks and \lam from
0-20\% and 20-60\% \cucu collisions at \sqrtsNN = 200 GeV. The
participant eccentricity $\varepsilon_{part}$ are from a Monte Carlo
Glauber calculation. (See Table~\ref{tab:glauber} for
$\varepsilon_{part}$.) After the geometric effect has been removed
by dividing by $\varepsilon_{part}$, the build-up of stronger
collective motion in more central collisions becomes obvious in the
measured elliptic flow.

Number of quark scaling was observed in 200 GeV \auau collisions
firstly. In Figure~\ref{fig3}, we test $n_q$ scaling in 200 GeV
\cucu collisions. At low and intermediate $p_T$, scaling works well;
At high $p_T$, $v_2$ for \ks and \lam have large error bars, but are
consistent with $n_q$ fitted curve. We can draw the conclusion:
Number-of-Quark scaling was also observed in 200 GeV \cucu
collisions.

Hydrodynamic model can be used to calculate elliptic flow in heavy
ion collisions, preliminary ideal hydrodynamic model results are
from Pasi Huovinen. In Figure~\ref{fig4}, we compare experimental
data to ideal hydrodynamic model results in different centrality
bins. In central collisions, ideal hydrodynamic model under-predicts
$v_2$; in peripheral collisions, ideal hydrodynamic model
over-predicts $v_2$.

In order to study the system size dependence of scaling behavior, we
normalize the $n_q$-scaled elliptic flow($v_2$) by the participant
for different systems. (See Table~\ref{tab:glauber} for
$\varepsilon_{part}$.) Figure~\ref{fig5} shows the doubly scaled
quantities from 200 GeV 0-60\% \cucu and 0-80\% \auau collisions.
After the geometric effect has been removed by dividing by
$\varepsilon_{part}$, the build up of stronger collective motion in
larger system becomes obvious, which is similar to the centrality
dependence in \cucu and \auau collisions~\cite{yanlvetal} at
\sqrtsNN = 200 GeV.  If hydrodynamic limit has been reached,
$v_2/\varepsilon_{part}$ should be a constant for \cucu and \auau
collisions~\cite{thermalization}. This indicates that hydrodynamic
limit has not been saturated in \cucu collisions.

\section{Summary}
\label{concl} We present STAR preliminary results of $v_2$ for \ks,
\lam and \xi from 200 GeV \cucu collisions at RHIC. In order to
reduce non-flow effects, FTPC tracks have been used to estimate the
event plane. At low $p_T$, $v_2$ is found to be consistent with mass
ordering. Number-of-Quark scaling was also observed in 200 GeV \cucu
collisions at three centrality bins. Preliminary ideal Hydrodynamic
model results are used to compare with experimental data. It
under-predicts the elliptic flow in central collisions,
over-predicts the elliptic flow in peripheral collisions. Stronger
collective flow can be observed in the more central collisions or
the larger system. $v_2/\varepsilon_{part}$ is not a constant for
\cucu and \auau collisions. This indicates that hydrodynamic limit
has not been saturated in \cucu collisions.

\vskip \baselineskip

\section{Acknowledgments}
Many thanks to the organizers; many thanks to P. Huovinen for
discussions. The author is supported in part by NSFC under project
10775058 and MOE of China under project IRT0624.

\vfill\eject

\begin{thebibliography}{100}

\bibitem{NPA757} J. Adams, et al., (STAR Collaboration),Nucl. Phys. {\bf A757}, 102
 (2005).

\bibitem{PRL99} B.I. Abelev, et al., (STAR Collaboration), Phys. Rev. Lett.,
 {\bf99}, 112301(2007).

 \bibitem{flow2} J. Adams et al., (STAR Collaboration), Phys. Rev. Lett. {\bf 93}, 252301 (2004).

\bibitem{flow3} J. Adams et al., (STAR Collaboration), Phys. Rev. {\bf C 72}, 014904 (2005).

\bibitem{flow4} B.B. Back et al., (PHOBOS Collaboration), Phys. Rev. {\bf C 72}, 051901 (2005).

\bibitem{flow5} B.B. Back et al., (PHOBOS Collaboration), Phys. Rev. Lett. {\bf 94}, 122303 (2005).


\bibitem{flow6} S.S. Adler et al., (PHENIX Collaboration), Phys. Rev. Lett. {\bf 94}, 232302 (2005).

\bibitem{klv2_130GeV} C.~Adler {\it et al.} (STAR Collaboration),
Phys. Rev. Lett. {\bf 89}, 132301 (2002).
\bibitem{starklv2} J. Adams {\it et al.} (STAR Collaboration),
Phys. Rev. Lett. {\bf 92}, 052302 (2004).
\bibitem{flowPRC} J. Adams \emph{et al.} (STAR Collaboration),
Phys. Rev. C \textbf{72}, 014904 (2005).
\bibitem{starwp} J. Adams {\it et al.} (STAR Collaboration),
Nucl. Phys. {\bf A757}, 102 (2005).
\bibitem{v2minv} N.~Borghini, J.-Y.~Ollitrault,
Phys. Rev. C {\bf 70}, 064905 (2004).
\bibitem{NCQ}    X. Dong et al. Phy. Lett. {\bf B597}, (2004)328



\bibitem{STARtpc} K.~H.~Ackermann {\it et al.} (STAR
Collaboration), Nucl. Instrum. Methods A {\bf 499}, 624 (2003).
\bibitem{yanlvetal}    B.I. Abelev et al. (STAR Collaboration),  nucl-ex/0801.3466
\bibitem{thermalization} S. Voloshin and A.M. Poskanzer, Phy. Lett. {\bf B414}, 27 (2000).


\end{thebibliography}
\end{document}